\documentclass[12pt,preprint]{aastex}

\usepackage{graphicx}

\slugcomment{ApJLett, in Press}

\shorttitle{The Return of an Old Friend}
\shortauthors{Jewitt, Ishiguro, Agarwal}

\begin{document}

\title{Large Particles in Active Asteroid P/2010 A2
 \footnote{
     The data presented herein were obtained at the W.M. Keck Observatory, which is operated as a scientific partnership among the California Institute of Technology, the University of California and the National Aeronautics and Space Administration. The Observatory was made possible by the generous financial support of the W.M. Keck Foundation. 
}
}

\author{David Jewitt$^2$,
 Masateru Ishiguro$^3$ and
  Jessica Agarwal$^4$}
\affil{$^2$Department of Earth and Space Sciences and Department of Physics and Astronomy,
University of California at Los Angeles, 
595 Charles Young Drive East, \\
Los Angeles, CA 90095-1567\\
$^3$ Department of Physics and Astronomy, Seoul National University,
San 56-1 Shillimdong, Gwanak-gu, KR Seoul 151-742, Republic of Korea\\
$^4$ Max Planck Institute for Solar System Research, Max-Planck-Str. 2,
37191 Katlenburg-Lindau, Germany}

\email{jewitt@ucla.edu}

\begin{abstract}
Previously unknown asteroid P/2010 A2 rose to prominence in 2010 by forming a transient, comet-like tail consisting of ejected dust. The observed dust production was interpreted as either the result of a hypervelocity impact with a smaller body or of a rotational disruption.  We have re-observed this object, finding that large particles remain a full orbital period after the initial outburst.  In the intervening years, particles smaller than $\sim$3 mm in radius have been dispersed by radiation pressure, leaving only larger particles in the trail.  Since the total mass is dominated by the largest particles, the radiation pressure filtering allows us to obtain a more reliable estimate of the debris mass than was previously possible.  We find that the mass contained in the debris  is $\sim$5$\times$10$^8$ kg (assumed density 3000 kg m$^{-3}$), the ratio of the total debris mass to the nucleus mass is $\sim$0.1 and that events like P/2010 A2 contribute $<$3\% to the Zodiacal dust production rate.  Physical properties of the nucleus and debris are also determined.

\end{abstract}


\section{Introduction}
Object P/2010 A2 was first reported as a short-period comet in data taken UT 2010 January 6 (Kadota et al.~2010).  However, the  orbit is that of a main-belt asteroid (semi-major axis 2.290 AU, eccentricity 0.124 and inclination 5.3\degr~and Tisserand parameter with respect to Jupiter $T_J$ = 3.582), leading to its classification as an active asteroid (equivalently, a ``main-belt comet''; Hsieh and Jewitt 2006, Jewitt 2012).  Hubble Space Telescope images taken from 2010 January 25 to May 29, inclusive, show a structured, parallel-sided dust tail that is detached from a $\sim$120 m diameter parent nucleus (Jewitt et al.~2010).   Dynamical models taking into account both solar gravity and radiation pressure reveal that mass was impulsively ejected from the parent in 2009 March, consistent with an origin through hypervelocity impact or through mass-shedding as a result of rotational instability (Jewitt et al.~2010). Independent studies support both the impulsive nature and the timing of the ejection from P/2010 A2 (Snodgrass et al.~2010, Hainaut et al.~2012, Kleyna et al.~2012, but see Moreno et al.~2010). Pre-discovery observations from as early as UT 2009 November 22 were subsequently identified (Jewitt et al.~2011).  P/2010 A2 went undetected for its first 8 months as a result of its small solar elongation and incomplete coverage of the night sky by on-going sky surveys.

Here, we report new observations of the nucleus and dust taken on UT 2012 October 14, a full orbit after the inferred date of initial activity.

\section{Observations} 

We used the 10-meter diameter Keck I telescope located atop Mauna Kea, Hawaii.  Images were secured using the Low Resolution Imaging Spectrometer (LRIS) camera (Oke et al.~1995). The LRIS camera has two channels housing red and blue optimized charge-coupled devices and separated by a dichroic filter (we used the ``460'' dichroic, which has 50\% transmission at 4875\AA).  On the blue side we used a broadband B filter (center wavelength $\lambda_c$ = 4369\AA, full width at half maximum (FWHM) $\Delta \lambda$ = 880\AA) and on the red side an R filter ($\lambda_c$ = 6417\AA, $\Delta \lambda$ = 1185\AA).  All observations used the facility atmospheric dispersion compensator to correct for differential refraction, and the telescope was tracked at non-sidereal rates while autoguiding on fixed stars.   The image scale on both cameras was 0.135\arcsec~pixel$^{-1}$ and the useful field of view approximately 320\arcsec$\times$440\arcsec.  Atmospheric seeing was about 1.0\arcsec~FWHM and the sky above Mauna Kea was photometric.

The heliocentric and geocentric distances were 2.190 AU and 1.203 AU, respectively, and the phase angle 5.1\degr.   P/2010 A2 was identified at the telescope by its motion and by its distinctive morphology. We secured nine images simultaneously in the B and R filters, with exposures of 660 s and 600 s, respectively.  At the time of observation, the non-sidereal rates of the target were 38\arcsec~hr$^{-1}$ W and 13\arcsec~hr$^{-1}$ S, so that field stars and galaxies trailed in individual images by about 7\arcsec.  Photometric calibration was secured through observations of standard stars from Landolt (1992).

\section{Discussion}
The appearance of P/2010 A2 is shown in Figure \ref{image}.  A point-like object, presumed to be nucleus ``N'' from Jewitt et al.~(2010), appears embedded in a long, thin debris trail of FWHM 3.0$\pm$0.2\arcsec (2600$\pm$180 km), with extensions to the edge of the field of view both to the east and the west of the nucleus.  Close inspection shows that ``N'' is located near the northern edge of the debris trail, consistent with the southerly (-3.8\degr) latitude of the Earth relative to the orbit plane at the time of observation.  None of the filamentary ``X''-shaped trail structures observed in 2010 are apparent in the Keck data. 

\subsection{Nucleus Photometry} 

Subtraction of light from the near-nucleus dust trail limits the accuracy of nucleus photometry.  Fortunately, the trail surface brightness is well-behaved near the nucleus, allowing useful results to be obtained.  Following experimentation with different methods, we measured the nucleus using a 1.0\arcsec~radius circular aperture with background subtraction determined from the median signal in a contiguous annulus 1.3\arcsec~in outer radius.  Other methods of background subtraction gave comparable results.  The resulting apparent magnitudes are listed in Table (\ref{photometry}).   Estimated absolute uncertainties on the nucleus photometry are $\sim$0.1 mag. (in bother filters), which is small compared to the uncertainty introduced on the absolute magnitude by the unknown phase function correction.

A direct measure of the nucleus color is obtained by comparing the simultaneous R and B filter photometry.  We find B-R =  1.36$\pm$0.04, which is slightly redder than the Sun (for which B-R = 0.99, Hardorp 1982, Hartmann et al.~1990). The quoted uncertainty on B-R is the standard error from measurements of 8 images (we rejected one image in each filter because of overlap with a field star).  From examination of the data, we believe that the color could be larger or smaller by  $\sim$0.1 mag.~as a result of systematic errors, but we are confident that the nucleus is significantly redder than sunlight.  

Next, the apparent magnitudes were converted to absolute magnitudes (i.e.~scaled to unit heliocentric, geocentric distances and at 0\degr~phase angle) using

\begin{equation}
H_R = m_R - 5\log_{10}\left(R_{au} \Delta_{au} \right) + 2.5\log_{10}(\Phi(\alpha))
\label{inverse}
\end{equation}

\noindent in which $R_{au}$ and $\Delta_{au}$ are the heliocentric and geocentric distances, respectively, both expressed in AU, and $\Phi(\alpha)$ is the ratio of the brightness at phase angle $\alpha$ to that at phase angle 0\degr.   We employed the HG formalism (Bowell et al.~1989) with scattering parameter $g$ = 0.25, as appropriate for an S-type asteroid, consistent with the measured B-R color.
With these assumptions, 2.5$\log_{10}$($\Phi(5.1\degr)$) = -0.37 mag., and the total correction from apparent to absolute magnitudes is $m_R - H_R$ = 2.47 mag. by Equation (\ref{inverse}).  The correction is uncertain by several $\times$0.1 mag.~because we do not know the nucleus phase function.

We find $H_R$ = 21.41$\pm$0.03 (only the statistical uncertainty is quoted) whereas $H_V$ = 22.00$\pm$0.07 (standard error on the mean of 8 measurements) was reported by Jewitt et al.~(2010).  The difference, $H_V$ - $H_R$ = 0.59$\pm$0.08 is again redder than the solar color ($m_V - m_R$ = 0.35 in the Kron-Cousins system), but may be affected by differences in the near-nucleus dust environment between 2010 and 2012.  

The absolute magnitudes are related to the cross-section and albedo by 

\begin{equation}
C = 2.24\times10^{22} \pi p_{R}^{-1} 10^{0.4(m_{\odot}(R) - H_{R})} 
\label{geometric}
\end{equation}

\noindent in which $m_{\odot}(R)$ = -27.11 is the apparent red magnitude of the Sun and $p_{R}$ is the geometric albedo, which we assume to be 0.15 following Jewitt et al.~2010.  (Equation (\ref{geometric}) is given for R filter photometry; the equivalent relation for the B filter has $m_{\odot}(B)$ = -26.12).

The resulting nucleus cross-section is $C_n$ = 0.019 km$^2$, corresponding to an equal-area circle of radius $r_n$ = ($C_n/\pi)^{1/2}$ = 78 m.  The statistical errors on $C_n$ and $r_n$ are meaninglessly small ($\pm$3\% and $\pm$1.5\%, respectively) compared to the uncertainty introduced by the unknown albedo.  The latter is perhaps $\pm$50\%, translating to a $\pm$25\% uncertainty on the nucleus radius, or $r_n$ = 78$\pm$20 m, which we take as our best estimate.  With assumed density $\rho$ = 3000 kg m$^{-3}$ and a spherical shape, the nucleus mass is 3$\times$10$^9$ kg $\le M_n \le$ 11$\times$10$^9$ kg  and the gravitational escape velocity (neglecting rotation) is 0.07 $\le V_e \le$ 0.13 m s$^{-1}$.

\subsection{Debris Trail Photometry}
To determine the surface brightness profile we rotated the image to align the trail with the x-axis and interpolated the sky brightness determined from regions 4\arcsec to 12\arcsec~above and below the trail (Figure \ref{sb}).  Bumps in the profile result from imperfectly removed field stars and galaxies.  

We measured the color of the dust within a 5.4\arcsec~long segment centered on the nucleus having high surface brightness and being free from the contaminating effects of background sources. We find B-R = 1.07$\pm$0.06 (Table \ref{photometry}), which is less red than the nucleus but consistent with the color of the Sun.   The difference in color between the nucleus and the dust is significant. While the nearly neutral dust colors are broadly consistent with a C-type asteroid spectrum, the nucleus more closely resembles a redder, possibly S-type asteroid.   Indeed, S-types are common in the inner regions of the asteroid belt, so that the reddish nucleus color is not surprising.  The red colors can be produced by a high abundance of nano-phase iron particles from space weathering at the surface, as observed on asteroid (25143) Itokawa (Noguchi et al.~2011).   Hence it is possible that fresh debris from an S-type asteroid might be less red than the parent body as a result of the absence of space weathering on previously buried material, as noted by Kim et al.~(2012).  Our data are compatible with this conjecture, although it is perhaps surprising that fallback debris has not coated the surface with fresh (neutral) material.

Integrated R-filter photometry was obtained from the whole visible region of the trail, in a rectangular box 8\arcsec$\times$288\arcsec~(Table \ref{photometry}).  Using Equation (\ref{geometric}) and the same ($p_R$ = 0.15) albedo as the nucleus, the trail brightness corresponds to a scattering cross-section $C_e$ = 4.9 km$^2$ in dust.

\subsection{Dust Dynamical Models}

We computed models to follow the motion of spherical particles (density $\rho$ = 3000 kg m$^{-3}$) under the action of solar gravity and radiation pressure.  In these models, the ratio of the radiation pressure acceleration to gravitational acceleration is $\beta$ = 0.2 $a_{\mu}^{-1}$, where $a_{\mu}$ is the particle radius in microns.  A given position along the trail can be reached by different combinations of dust ejection velocity and $\beta$, such that we can determine a relation between these parameters at a given nucleus distance.
To study this relation, we did not take into account the velocity component perpendicular to the orbital plane of the nucleus, which merely influences the width of the trail.

We calculated the positions of $4 \times 10^6$ test particles ejected on UT 2009 March 02 parallel to the orbital plane of the nucleus and having $0 \leq \beta \leq 10^{-5}$ in steps of $10^{-7}$. The velocities range from -1 m s$^{-1}$ to +1 m s$^{-1}$ in steps of 0.01 m s$^{-1}$, independently in both $v_x$ and $v_y$, where $v_y$ is parallel to the orbital motion of the nucleus at the time of ejection, and $v_x$ is perpendicular to it, pointing away from the Sun.  For any given distance from the nucleus $x$, we find a linear relationship between $v_y$ and $\beta$, while there is no significant correlation with $v_x$. We also find a linear relation between the nucleus distance and $v_y (\beta=0)$: 

\begin{equation}
\label{fit}
\beta(v_y, x) = c_1 v_y - c_2 x,
\end{equation}

\noindent where $v_y$ is in m s$^{-1}$, and $x$ is the distance from the nucleus in RA in arcsec (positive $x$ corresponding to location east of the nucleus). The numerical values of the constants are $c_1 = -7.5 \times 10^{-5}$ and $c_2 = 2.2 \times 10^{-7}$.   

By Equation (\ref{fit}), material projected to the east of the nucleus (leading the nucleus in its orbital motion) must have been ejected opposite to the direction of motion of the nucleus.
Material west of the nucleus could have been ejected in either direction, depending on its $\beta$-parameter. 

From the extrapolated eastern extension of the nucleus ($x=50\arcsec \pm 10\arcsec$, cf. Fig.~\ref{sb}), we find that particles must have been ejected with $v \leq -0.15 \pm 0.03$ m s$^{-1}$.  The smallest particles to the east of the nucleus are those pushed back to $x$ = 0 by radiation pressure.  By Equation ({\ref{fit}) these have $\beta$ = 1.1$\times$10$^{-5}$ for $v=-0.15$ m s$^{-1}$, corresponding to $a$ = 1.8 cm.  We infer that sub-centimeter particles have been entirely swept from the east arm of the trail.

At the western edge of the field of view ($x$ = -250\arcsec), we find for $v_y=-0.15~{\rm m~s^{-1}}$ particles having 
$\beta = 7 \times 10^{-5}$ (radius  3\,mm). Effectively, exposure to the Sun over one orbit period has acted as a filter to remove smaller particles from the dust size distribution. 

We used a 3D numerical model to confirm and amplify these results, again assuming isotropic ejection on 2009 March 2.   
The model assumes that the dust particle radii are distributed as a power law, $dN(a) = \Gamma a^{-q} da$, where $dN(a)$ is the number of particles having radii in the range $a$ to $a + da$, and $\Gamma$ and $q$ are constants.  We used the model to explore the allowable values of the minimum and maximum particle radii, $a_{min}$ and $a_{max}$, respectively, the ejection speed, $v$, and $q$.  

The model (Figure \ref{sb2}) shows that the slope of the surface brightness profile west of the nucleus is primarily a function of $q$, with acceptable fits requiring $q$ = 3.5$\pm$0.1.  Earlier determinations include $q$ = 3.3$\pm$0.2 (Jewitt et al.~2010), $q$ = 3.4$\pm$0.3 (Moreno et al.~2010), $q$ = 3.5 (no uncertainty quoted: Snodgrass et al.~2010), $q$ = 3.44$\pm$0.08 (Hainaut et al.~2012), all of which are consistent with the value determined here.  For comparison, ejecta from the impact of the Deep Impact spacecraft into the nucleus of comet P/Tempel 1 had $q$ = 3.1$\pm$0.3, but $q$ is only well-determined for particles with $a <$ 20 $\mu$m (Jorda et al.~2007). 

The shape and extent of the profile east of the nucleus are controlled mainly by the largest particles and their ejection speed.   With $q$ = 3.5, we find a best fit with $\beta_{min}$ = 1$\times$10$^{-6}$ ($a_{max}$ = 0.2 m) and $v$ = 0.15 m s$^{-1}$.  Uncertainties on these values are approximately a factor of two, within the context of the model. 

The ``X''-shaped structure observed in HST data from 2010 is not seen in our Keck images.  Grain motions perpendicular to the orbital plane are unaffected by radiation pressure, causing the vertical extent of the ``X''  to vary sinusoidally with time. Based on the HST data, we expect that the cross in Figure (\ref{image}) had a vertical extent of 0.9\,arcsec, which is below our seeing limit.

\subsection{Mass}
With the above parameters, we find that the cross-section weighted mean particle size is $\overline{a} = (a_{min} a_{max})^{1/2} \sim$ 0.02 m.  For uniform spheres, the mass, $M_d$, and the effective cross-section, $C_e$, are related by 

\begin{equation}
M_d = (4/3) \rho \overline{a} C_e, 
\label{mass}
\end{equation}

\noindent where $\rho$ is the density of the grains.  Substituting $C_e$ = 4.9 km$^2$, we obtain $M_d$ = 4$\times$10$^8$ kg for the mass in particles with 3 mm $\le a \le$ 0.2 m in the Keck field of view.

We next estimated a correction for the fraction of the trail missing from the Keck field of view, in two ways.  First, we fitted a smooth function to the linearized surface brightness profile (Figure \ref{sb}) and extrapolated the function until the surface brightness reached zero at either end of the trail.  We found that $\sim$12\% of the scattering cross-section fell outside the Keck field of view.  Including this correction and integrating over the size distribution from 1 $\mu$m to 0.2 m gives an estimated total mass 5$\times$10$^8$ kg, with an uncertainty of at least a factor of two.  Second, we integrated the fitted 3D model over the size distribution and beyond the limits of the CCD field of view, finding a total mass 
(5 to 6)$\times$10$^8$ kg.  If the maximum particle radius is $>$0.2 m, then the total mass would be larger still (in proportion to $a_{max}^{1/2}$). Dust masses ((0.6 to 6)$\times$10$^8$ kg by Jewitt et al.~2010, 8$\times$10$^8$ kg by Hainaut et al.~2012) were inferred from photometry obtained in 2010.

The fate of the particles in Figure (\ref{image}) is to be collisionally shattered (collisional lifetime of centimeter-sized grains is $\le$10$^4$ yr  (Grun et al.~1985)), contributing debris to  the  Zodiacal complex.  Events like P/2010 A2 are thought to occur perhaps twice per year, although only $\sim$6\% are detected by current surveys (Jewitt et al.~2011).  The resulting average mass production rate is $\sim$30 kg s$^{-1}$.  For comparison, estimates of the dust production rate needed to maintain the Zodiacal cloud in steady-state range from $\sim$10$^3$ kg s$^{-1}$ (Leinert et al.~1983) to 10$^4$ or even 10$^5$ kg s$^{-1}$ (Nesvorny et al.~2011).   We conclude that $<$3\% of the Zodiacal dust production rate is from events like P/2010 A2, and that Zodiacal dust production must be dominated by another source.

The ratio of the debris mass to the nucleus mass (which is independent of the assumed albedo provided the nucleus and dust have the same albedo) is $\frac{M_d}{M_n} \sim$ 0.1.  The uncertainty on this ratio is at least a factor of two, resulting from assumptions about the shape of the nucleus, the density, and the upper limit to the size of the ejected boulders.  Evidently, the mass in the trail is a substantial fraction of the mass in the nucleus.

Interpreted as an impact, we can relate the ejecta mass to the projectile mass as follows.  At an assumed impact speed $\Delta V$ = 5$\times$10$^3$ m s$^{-1}$ and with escape velocity $V_e \sim$ 0.1 m s$^{-1}$, the ratio of ejecta mass to projectile mass is $f$ = $M_d/m_p \sim$ 10$^4$ (Housen and Holsapple 2011).  If the projectile and the nucleus of P/2010 A2 are of equal density, then substitution into Equation (\ref{mass}) gives the projectile radius

\begin{equation}
r_p = \left(\frac{\overline{a} C_e}{\pi f}\right)^{1/3}.
\label{projectile}
\end{equation}

Equation (\ref{projectile}) gives $r_p$ = 1.5 m,  and a specific energy $E/M_n$ = (1/2)($r_p/r_n)^3 \Delta V^2 \sim$ 10$^2$ J kg$^{-1}$,  comparable to the value needed for catastrophic disruption (Jutzi et al.~2010).  Simulations of catastrophically disrupted sub-kilometer sized bodies yield fragments with differential size distribution indices -3.2 $\le q \le$ -3.7 and median ejection speeds $\sim$0.1 m s$^{-1}$ (Jutzi et al.~2010), both similar to values measured in P/2010 A2 .  We conclude that the data are consistent with P/2010 A2 being caused by an impact close to the disruption threshold. 

Separately, the properties of P/2010 A2 remain consistent with those expected of a rotationally disrupted body (as noted in Jewitt et al.~2010), as far as these are known (Marzari et al.~2011, Jacobsen and Scheeres 2011).  The ejecta/nucleus mass ratio, $M_d/M_n \sim$ 0.1, the impulsive nature of the mass loss and the very low velocity of the debris all fit qualitatively with expectations from rotational break-up.  The rotation of the nucleus offers a possible observational discriminant between the impact and rotational mass-shedding hypotheses, with rotational disruption requiring rapid spin.    Unfortunately, an attempt to detect the rotational lightcurve in the nucleus from Keck data, albeit from a very short ($\sim$1 hr) data-arc, was unsuccessful owing to the extreme faintness of this body.

\section{Summary}

From new observations of active asteroid P/2010 A2 we find that

\begin{enumerate}

\item The nucleus has  absolute magnitude $H_R$ = 21.41$\pm$0.03 and is redder than the surrounding dust (B-R = 1.36$\pm$0.04 vs.~B-R = 1.07$\pm$0.06).  These measurements suggest the excavation of unreddened (C-type) material from beneath the surface of a space-weathered (S-type) asteroid of  radius $r_n$ = 78$(0.15/p_R)^{1/2}$ m, where $p_R$ is the unmeasured red geometric albedo. 

 \item Large dust particles (radii 3 mm $\le a \le$ 20 cm, differential size distribution index $q$ = 3.5,  effective mean radius near 2 cm) persist near the nucleus a full orbit after their impulsive release in 2009.  The sum of the scattering cross-sections of the dust is $\sim$5 km$^2$, corresponding to mass $\sim$5$\times$10$^8$ kg, or about 10\% of the mass of the surviving nucleus. 

\item Events comparable to P/2010 A2 contribute $\lesssim$3\% to the source of dust for the Zodiacal cloud complex.

\item The known properties of P/2010 A2 do not allow us to distinguish between impact and rotational disruption origins.

\end{enumerate}

\acknowledgments
We thank Luca Ricci (LRIS) and Julie Renaud-Kim (Keck) for assistance, Jing Li and the anonymous referee for comments. This work was supported by a grant to DCJ from NASA's Planetary Astronomy program.

\clearpage


%



%

\clearpage

\begin{deluxetable}{llllccc}
\tablecaption{Photometry
\label{photometry}}
\tablewidth{0pt}
\tablehead{
\colhead{Feature}  & \colhead{Filter}  & \colhead{Region [$\arcsec$]\tablenotemark{a}} & \colhead{Size [10$^3$ km] \tablenotemark{b}}   & \colhead{$m_R$ [deg]\tablenotemark{c}} & \colhead{$H$  \tablenotemark{d}}  & \colhead{$C [km^2]$  \tablenotemark{e}} }
\startdata
Nucleus & R & Circle 1.0 & 0.88 radius & 23.88$\pm$0.03 & 21.41$\pm$0.03 & 0.019$\pm$0.009 \\
Nucleus & B & Circle 1.0 & 0.88 radius & 25.24$\pm$0.02 & 22.77$\pm$0.02 & 0.013$\pm$0.006 \\

Trail & R & Box 5.4$\times$5.4 & 4.74$\times$4.74 & 22.93$\pm$0.04 & 20.46$\pm$0.04 & 0.044$\pm$0.002 \\
Trail & B & Box 5.4$\times$5.4 & 4.74$\times$4.74 & 24.00$\pm$0.04 & 21.53$\pm$0.04 & 0.041$\pm$0.001 \\

Trail\tablenotemark{f} & R & Box 8$\times$288 & 7.03$\times$253 & 17.8 &  15.3  & 4.9\\


\enddata


\tablenotetext{a}{Angular dimensions of the region measured, in arcsec, with the long axis parallel to the trail}
\tablenotetext{b}{Corresponding linear dimensions of the region measured, in 10$^3$ km}
\tablenotetext{c}{Measured apparent magnitude}
\tablenotetext{d}{Absolute magnitude computed from Equation (\ref{inverse})}
\tablenotetext{e}{Equivalent scattering cross-section from Equation (\ref{geometric}), in km$^2$}
\tablenotetext{f}{Uncertainties in this large aperture measurement are dominated by systematics from the sky background subtraction}

\end{deluxetable}

\clearpage

\begin{figure}
\epsscale{1.00}
\begin{center}
\plotone{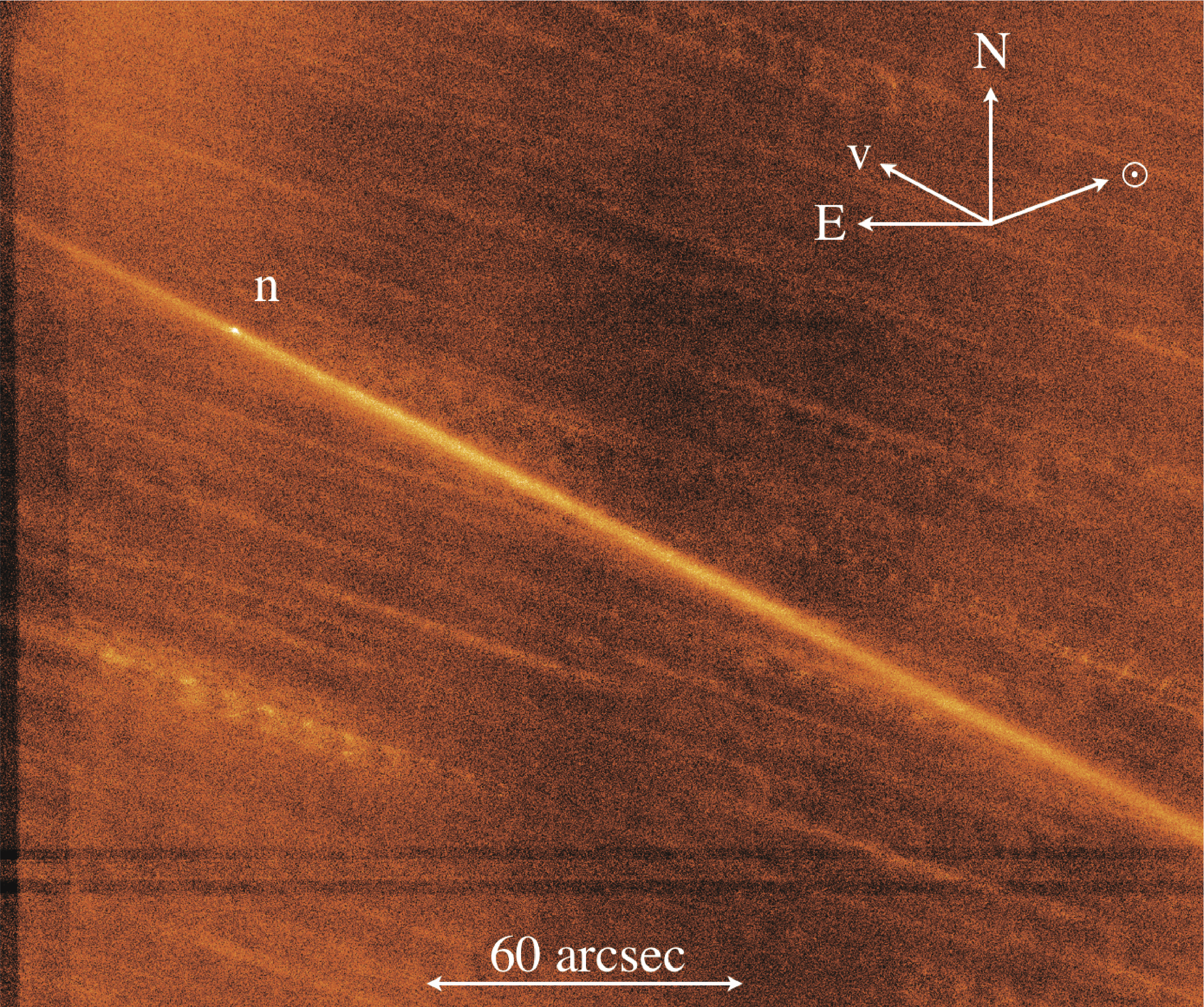}
\caption{Keck image of P/2010 A2 taken on UT 2012 October 14, computed from the median of nine, non-sidereally tracked R-filter images each of 600 s duration. The cardinal directions are marked, as are the direction to the Sun ($\odot$) and the projected heliocentric velocity vector ($V$).   The nucleus is marked ``n'' on the trail towards the North-East. The stippled background is due to imperfectly removed field galaxies.  \label{image}
} 
\end{center} 
\end{figure}

\clearpage

\begin{figure}
\epsscale{1.00}
\begin{center}
\plotone{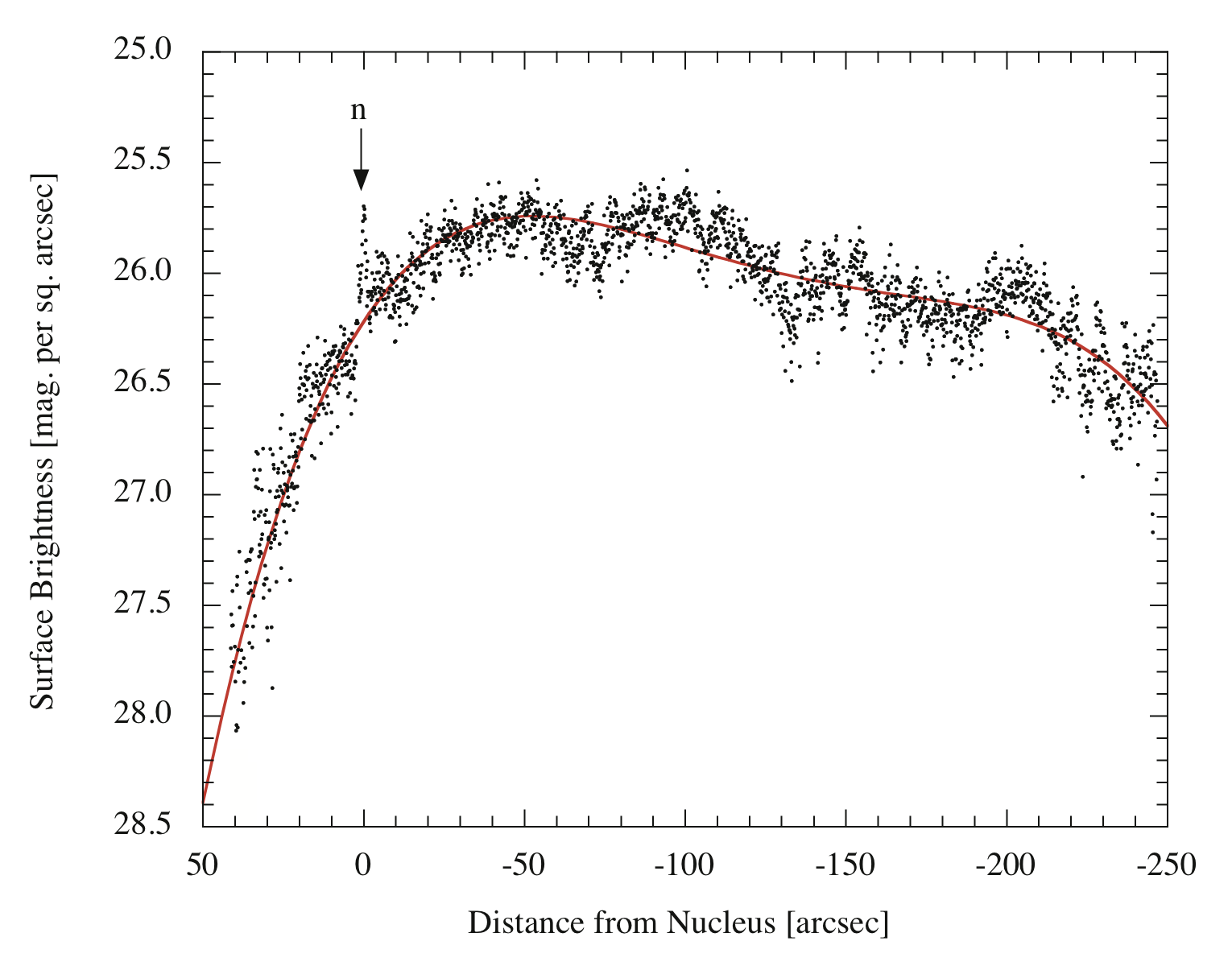}
\caption{The R-band surface brightness measured along the axis of the trail and averaged over a region 8.1\arcsec~perpendicular to the trail.  Distances are measured relative to the point-like nucleus ``n'', with positive distances being East.  The red line shows a smooth function fitted to the data to guide the eye.  The B-band surface brightness profile is indistinguishable within the measurement uncertainties. Bumps on the profile are background subtraction errors. \label{sb}
} 
\end{center} 
\end{figure}

\clearpage

\begin{figure}
\epsscale{1.00}
\begin{center}
\plotone{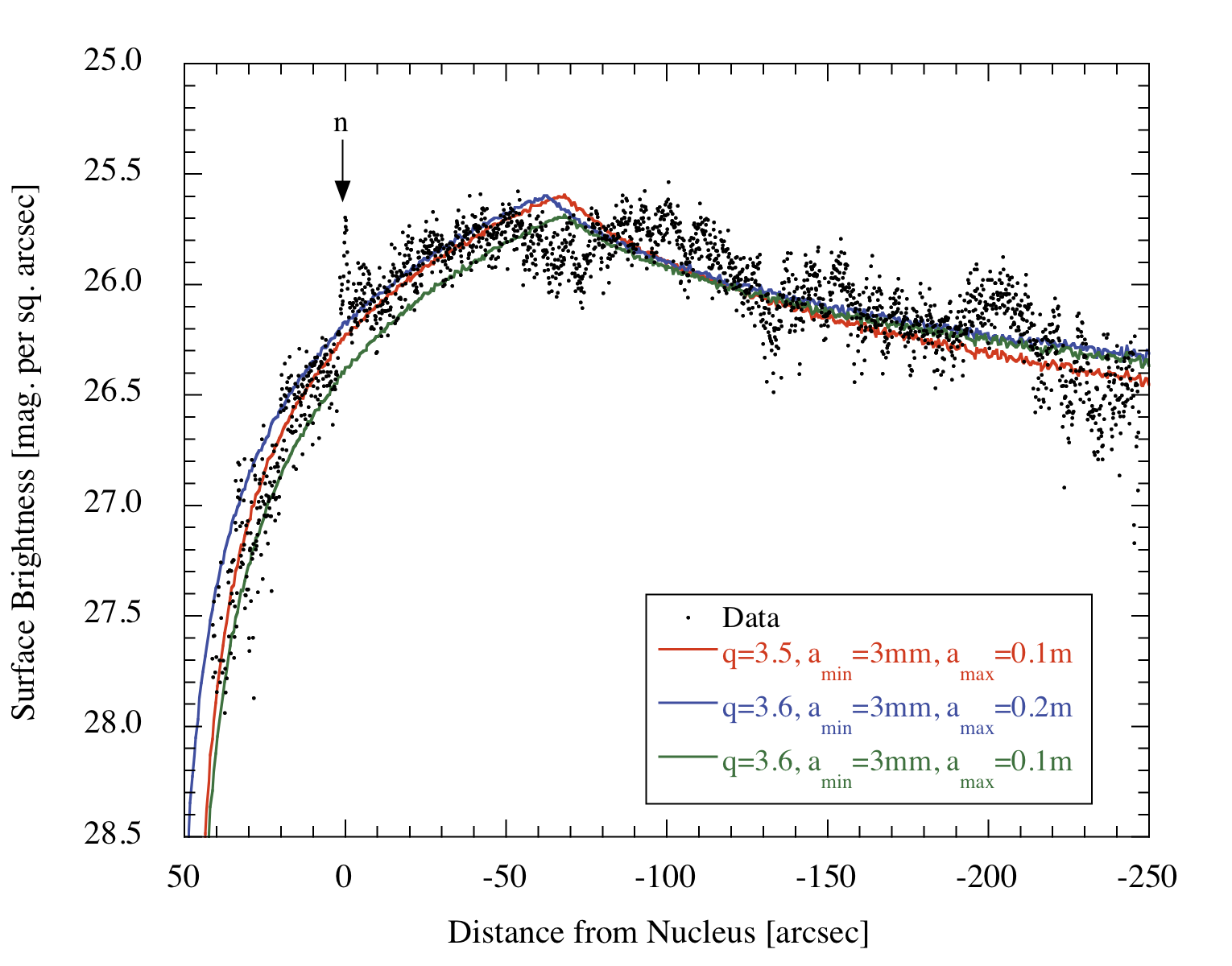}
\caption{Same as Figure \ref{sb} with sample models over-plotted.    \label{sb2}
} 
\end{center} 
\end{figure}

\clearpage

\begin{figure}
\epsscale{1.00}
\begin{center}
\plotone{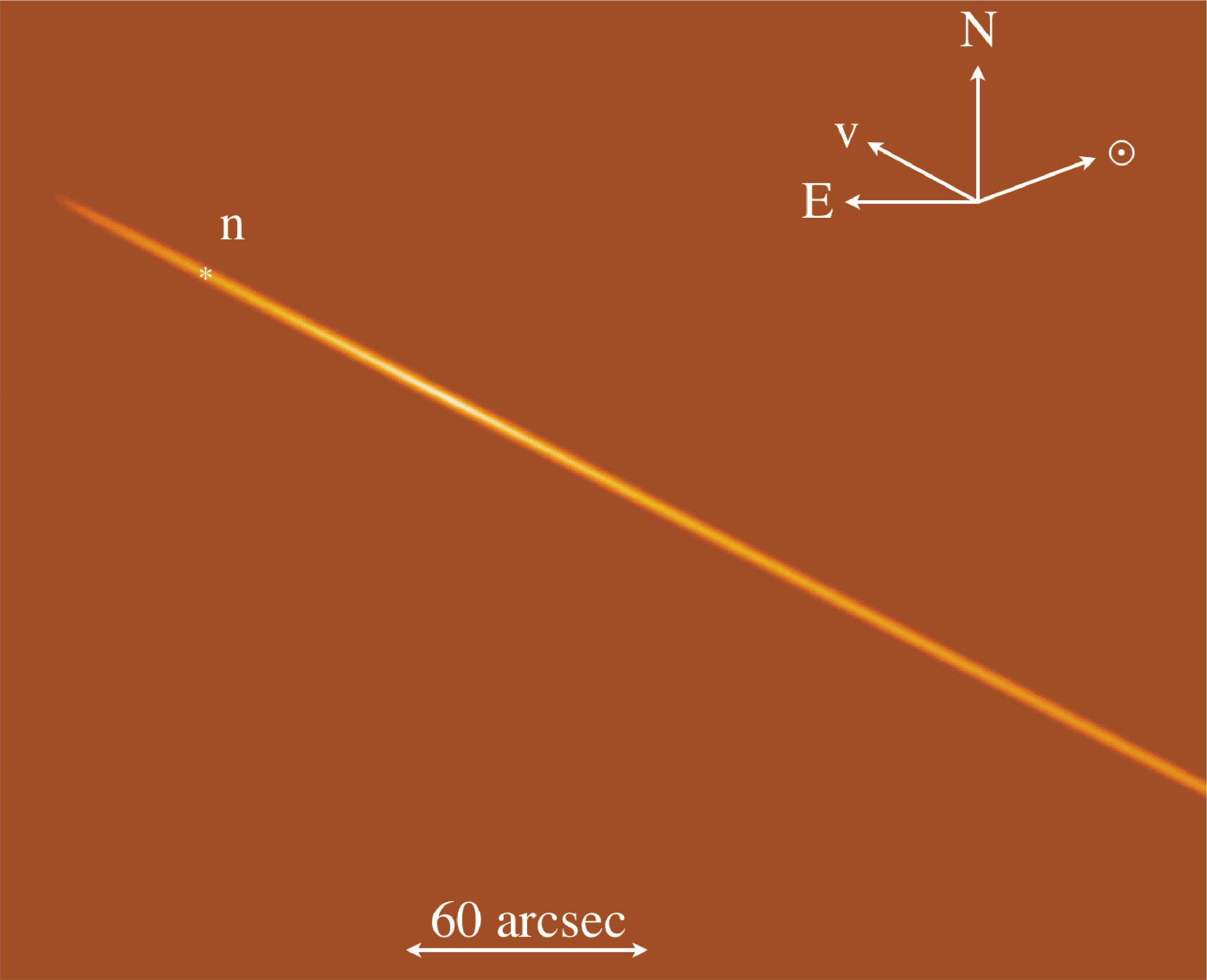}
\caption{Model of the dust trail with $q$ = 3.5, $a_{min}$ = 3 mm and $a_{max}$ = 20 cm, smoothed to match the 1\arcsec~FWHM of the data.  \label{model}
} 
\end{center} 
\end{figure}

\clearpage


\clearpage

\begin{thebibliography}{}

\bibitem[Bauer et al.(2012)]{2012ApJ...747...49B} Bauer, J.~M., Mainzer, 
A.~K., Grav, T., et al.\ 2012, \apj, 747, 49 

\bibitem[Benz 
\& Asphaug(1999)]{1999Icar..142....5B} Benz, W., \& Asphaug, E.\ 1999, Icarus, 142, 5 



\bibitem[Bowell et al.(1989)]{1989aste.conf..524B} Bowell, E., Hapke, B., 
Domingue, D., et al.\ 1989, Asteroids II, 524 

\bibitem[Finson 
\& Probstein(1968)]{1968ApJ...154..327F} Finson, M.~J., \& Probstein, R.~F.\ 1968, \apj, 154, 327 

\bibitem[Grun et al.(1985)]{1985Icar...62..244G} Grun, E., Zook, H.~A., 
Fechtig, H., \& Giese, R.~H.\ 1985, Icarus, 62, 244 


\bibitem[Hainaut et 
al.(2012)]{2012A&A...537A..69H} Hainaut, O.~R., Kleyna, J., Sarid, G., et al.\ 2012, \aap, 537, A69 

\bibitem[Hardorp(1982)]{1982A&A...105..120H} Hardorp, J.\ 1982, \aap, 105, 120 


\bibitem[Hartmann et al.(1990)]{1990Icar...83....1H} Hartmann, W.~K., 
Tholen, D.~J., Meech, K.~J., \& Cruikshank, D.~P.\ 1990, Icarus, 83, 1 

\bibitem[Housen 
\& Holsapple(2011)]{2011Icar..211..856H} Housen, K.~R., \& Holsapple, K.~A.\ 2011, Icarus, 211, 856 

\bibitem[Hsieh 
\& Jewitt(2006)]{2006Sci...312..561H} Hsieh, H.~H., \& Jewitt, D.\ 2006, Science, 312, 561 


\bibitem[Jacobson 
\& Scheeres(2011)]{2011Icar..214..161J} Jacobson, S.~A., \& Scheeres, D.~J.\ 2011, Icarus, 214, 161 


\bibitem[Jewitt(2012)]{2012AJ....143...66J} Jewitt, D.\ 2012, \aj, 143, 66 

\bibitem[Jewitt et al.(2010)]{2010Natur.467..817J} Jewitt, D., Weaver, H., 
Agarwal, J., Mutchler, M., \& Drahus, M.\ 2010, \nat, 467, 817 

\bibitem[Jewitt et al.(2011)]{2011AJ....142...28J} Jewitt, D., Stuart, 
J.~S., \& Li, J.\ 2011, \aj, 142, 28 

\bibitem[Jorda et al.(2007)]{2007Icar..187..208J} Jorda, L., Lamy, P., 
Faury, G., et al.\ 2007, Icarus, 187, 208 

\bibitem[Jutzi et al.(2010)]{2010Icar..207...54J} Jutzi, M., Michel, P., 
Benz, W., \& Richardson, D.~C.\ 2010, Icarus, 207, 54 


\bibitem[Kadota et al.(2010)]{2010MPEC....A...32K} Kadota, K., Blythe, M., 
Spitz, G., et al.\ 2010, Minor Planet Electronic Circulars, 32 


\bibitem[Kim et al.(2012)]{2012ApJ...746L..11K} Kim, J., Ishiguro, M., 
Hanayama, H., et al.\ 2012, \apjl, 746, L11 

\bibitem[Kleyna et al.(2012)]{2012arXiv1209.2210K} Kleyna, J., Hainaut, O., 
\& Meech, K.\ 2012, arXiv:1209.2210 

\bibitem[Landolt(1992)]{1992AJ....104..340L} Landolt, A.~U.\ 1992, \aj, 
104, 340 

\bibitem[Leinert et 
al.(1983)]{1983A&A...118..345L} Leinert, C., Roser, S., \& Buitrago, J.\ 1983, \aap, 118, 345 

\bibitem[Marzari et al.(2011)]{2011Icar..214..622M} Marzari, F., Rossi, A., 
\& Scheeres, D.~J.\ 2011, Icarus, 214, 622 

\bibitem[Moreno et al.(2010)]{2010ApJ...718L.132M} Moreno, F., Licandro, 
J., Tozzi, G.-P., et al.\ 2010, \apjl, 718, L132 

\bibitem[Nesvorn{\'y} et al.(2011)]{2011ApJ...743..129N} Nesvorn{\'y}, D., 
Janches, D., Vokrouhlick{\'y}, D., et al.\ 2011, \apj, 743, 129 


\bibitem[Noguchi et al.(2011)]{2011Sci...333.1121N} Noguchi, T., Nakamura, 
T., Kimura, M., et al.\ 2011, Science, 333, 1121 



\bibitem[Oke et al.(1995)]{1995PASP..107..375O} Oke, J.~B., Cohen, J.~G., 
Carr, M., et al.\ 1995, \pasp, 107, 375 

\bibitem[Snodgrass et al.(2010)]{2010Natur.467..814S} Snodgrass, C., 
Tubiana, C., Vincent, J.-B., et al.\ 2010, \nat, 467, 814 


\end{thebibliography}
\end{document}